\documentstyle[twocolumn,prl,aps]{revtex}
\begin{document}
\twocolumn[\hsize\textwidth\columnwidth\hsize\csname @twocolumnfalse\endcsname

\title{Electronic structure of NiS$_{1-x}$Se$_x$
across the phase transition}
\author{D. D. Sarma \cite{jnc}, S. R. Krishnakumar, and Nirmala
Chandrasekharan}
\address{Solid State and Structural Chemistry Unit, Indian Institute
of Science, Bangalore 560012, India}
\author{E. Weschke, C. Sch\"u{\ss}ler-Langeheine, L. Kilian and G. Kaindl}
\address{Institut f\"ur Experimentalphysik, Freie Universit\"at Berlin,
D-14195 Berlin-Dahlem, Germany}
\date{\today}
\maketitle

\begin{abstract}

We report very highly resolved photoemission spectra of
NiS$_{1-x}$Se$_x$ across the so-called metal-insulator transition as
a function of temperature as well as composition. The present results
convincingly demonstrate that the low temperature, antiferromagnetic
phase is metallic, with a reduced density of states at E$_F$. This
decrease is possibly due to the opening of gaps along specific
directions in the Brillouin zone caused by the antiferromagnetic
ordering.

\end{abstract}

]

        While metal-insulator transition has been one of the
outstanding problems in condensed matter physics~\cite{Mott1}, the
specific case of a transition in hexagonal NiS~\cite{Sparks1} as a
function of temperature has been possibly the most controversial,
spanning a period of three decades. The controversy relates to the
very basic issue of the nature of the transition, with various groups
describing it as a metal-insulator
transition~\cite{Sparks1,Sparks2,Sparks3,Townsend1,Barker1,Nakamura1},
while nearly equal number of groups claiming it to be a metal-metal
transition~\cite{Mott2,Mcwhan1,Koehler1,Coey1,Anzai1}. The transport
and magnetic properties are that NiS  is a highly conducting
($\approx$10$^{-5}$~ohm-cm) Pauli paramagnetic metal at room
temperature with a characteristic metallic dependence of resistivity
on temperature~\cite{Koehler1,Coey1}. Near 260~K, the system
undergoes a first order phase transition, with a nearly two orders of
magnitude increase in resistivity
($\approx$10$^{-3}$~ohm-cm)~\cite{Koehler1,Coey1}.  The system
becomes antiferromagnetic below the transition~\cite{Sparks2},
showing a volume expansion of about~$\approx$1.9\% without any change
in crystal symmetry.  While it is generally agreed that the high
temperature phase represents an example of a highly conducting
metallic compound, the controversy continues to exist concerning the
nature of the low temperature phase.  Normally one would expect that
transport measurements would readily resolve the question concerning
the metallic or insulating phase of the ground state. A very unusual
nearly temperature independent resistivity down to about
4~K~\cite{Koehler1,Coey1}, however, does not provide any unambiguous
clue.  While in a metallic system the resistivity is expected to
decrease with temperature, in contrast to the observed behavior, an
insulating ground state should have very manifest exponentially
increasing resistivity, when the temperature scale is lower than the
band gap.  Thus, if the low temperature phase is insulating, the band
gap should be much smaller than 4~K~($<$ 0.5~meV). On the other hand,
if it is metallic, it has to be an unusual state where the
resistivity is nearly independent of temperature over a wide range of
temperature (4~K~$\leq$~T~$\leq$~260~K).  Unfortunately, theoretical
studies have not been of much help in settling these discussions in
favor of either of the views.  The earliest non-self-consistent band
structure calculations~\cite{Mattheiss1} correctly predicted the high
temperature Pauli paramagnetic phase to be metallic, while the low
temperature anti-ferromagnetic phase yielded a small gap. However,
subsequent self-consistent band structure
calculations~\cite{Fujimori1} based on local spin density
approximation~(LSDA) failed to yield any gap.  Similarly, the LDA+U
method~\cite{Anisimov1} obtained a metallic state for the low
temperature phase. In contrast, a recent improved version of the
LDA+U method has obtained a band gap for the low temperature
phase~\cite{Anisimov2}, which is by far too large to be consistent
with a temperature independent resistivity down to 4~K.  Thus,
neither transport measurements nor theoretical calculations have been
able to resolve the controversy surrounding the ground state behavior
of NiS.

        Metallic and insulating behavior are distinguished by the
presence or the absence of finite density of states (DOS) at the
Fermi energy, which can be probed by various spectroscopic
techniques. These have so far favored an insulating ground state for
NiS~\cite{Nakamura1}. An early optical reflectivity study showed the
existence of a characteristic dip at about 140~meV for the low
temperature phase compared to the high temperature metallic
phase~\cite{Barker1}; this has been cited~\cite{Nakamura1} as a proof
for the existence of a band gap of the same order.  Photoelectron
spectroscopy, which is the only technique that can directly probe the
density of states at E$_F$, also provided evidence for an insulating
state of NiS at low temperature. Specifically, a moderately high
resolution ($\Delta$E~$\approx$~36~meV) photoelectron spectrum of NiS
at low temperature has been reported~\cite{Nakamura1}. While it
exhibits finite spectral weight at E$_F$, the authors concluded on
the basis of an analysis of the density of states (DOS) and various
broadening effects such as resolution and thermal (Fermi-Dirac)
broadening that there is a gap of $\approx$~10~meV in the occupied
part of the DOS. The analysis also suggested a finite DOS at the band
edge, which is very unusual for a system with a three-dimensional
crystal structure. Since discontinuous band edges are only expected
in one and two dimensions, this finding is of special significance,
implying a novel consequence of electron
correlations~\cite{Nakamura1}.  Thus, the combination of various
spectroscopic results suggests that 1)~there is a total band gap of
about 140~meV in NiS, and 2)~the leading edge of the occupied DOS is
about 10~meV below E$_F$, with the remaining part (130~meV) of the
band gap presumably occurring in the unoccupied part. Such a
scenario, however, is in complete disagreement with the temperature
independent transport data, which suggest an upper limit of the band
gap, if any, to be less that 4~K~($<$~0.5~meV). In order to address
this unresolved puzzle concerning the nature of the transition and
the ground state of NiS, we have re-investigated the electronic
structures in NiS and several related compounds using
temperature-dependent photoelectron spectroscopy performed with very
high energy resolution. Our study conclusively shows that the ground
state is metallic and thus, the phase transition represents a
metal-to-{\it anomalous-metal} transition with decreasing
temperature.

      Temperature-dependent photoemission (PE) experiments were
carried out on polycrystalline samples of NiS$_{1-x}$Se$_x$, with
x=0.0, 0.11, 0.15, and 0.17, which were prepared by solid state
reaction in sealed quartz tubes~\cite{Anzai1}.  The samples were
characterized before and after the PE experiments by x-ray
diffraction. The samples with x=0.0 and x=0.11  showed a sharp
increase of the resistivity at transition temperatures T$_t$=260~K
and 97~K, respectively, while for x=0.15 and x=0.17 no transition was
observed~\cite{Sarma}. The samples were mounted on a Cu sample holder
fitted to a continuous-flow He cryostat allowing measurements between
25~K and 300~K. They were cleaned in situ by repeated scraping with a
diamond file; sample cleanness was checked by valence-band PE at
h${\nu}$=40.8~eV. PE spectra were recorded with a Scienta SES-200
electron energy analyzer using a Gammadata VUV-5000 photon source for
excitation.

        Addressing the question of the DOS close to E$_F$ in a PE
experiment imposes severe requirements as well on the experimental
resolution as on the stability of the experimental Fermi-level
position. In our experiment both were monitored by repeatedly
recording the Fermi edge region of polycrystalline Ag mounted close
to the NiS$_{1-x}$Se$_x$ samples. The total-system resolution
achieved in these experiments was 9$\pm1$~meV at h${\nu}$=21.2~eV, as
determined from the Ag Fermi edge at 25~K, see Fig.~1. Furthermore,
from the numerous Ag spectra, which were always recorded as reference
before and after taking data from the NiS$_{1-x}$Se$_x$ samples, we
found the position of E$_F$ to be reproducible within 1~meV. As
indicated in Fig.~1, even a shift as small as 2~meV would be clearly
visible. Another issue which has to be considered is the difficulty
in determining the Fermi level position from a PE spectrum in a
system where the DOS is not slowly varying across E$_F$. In order to
avoid such ambiguities, spectra were taken also from
NiS$_{0.85}$Se$_{0.15}$ and NiS$_{0.83}$Se$_{0.17}$ as reference
systems with very similar electronic structure as NiS and
NiS$_{0.89}$Se$_{0.11}$, which, however, do not undergo the phase
transition and stay metallic in the whole temperature range relevant
in this study. Together with NiS, Fig.~1 displays PE spectra of
NiS$_{0.83}$Se$_{0.17}$ taken at 25~K, and an  arbitrarily scaled Ag
spectrum. Evidently, the inflection points of all curves are at the
same energy. We have analyzed the three spectra in Fig.~1 in terms of
polynomial expressions for the DOS in order to determine the Fermi
energy in each case. It is found that the three independently
determined Fermi energies for these three systems coincide within
less than 1~meV. This result strongly suggests that the DOS of NiS is
metallic also in the low-temperature phase, which will be
demonstrated further in the following.

        Fig.~2 displays comparative PE spectra of all four
NiS$_{1-x}$Se$_x$ compounds taken above T$_t$ at 300~K (left panel),
and below T$_t$ (right panel). As shown in the insets, all spectra
are normalized to equal intensities at 680~meV binding energy (BE),
which leads to proper normalization over the whole spectral range at
higher BE's including the Ni~$d$-bands.  At T=300~K, the spectra of
all systems are identical in the narrow range around E$_F$,
reflecting identical metallic DOS.  Below T$_t$, distinct differences
are observed between those systems, which undergo the phase
transition (NiS, NiS$_{0.89}$Se$_{0.11}$) and those which do not
(NiS$_{0.85}$Se$_{0.15}$, NiS$_{0.83}$Se$_{0.17}$).  The data
analysis clearly shows that the spectra of the latter group 
($x$=0.15 and 0.17) are characterized by the same DOS both for the 300
and 25~K spectra; the same DOS is also consistent with the 300~K
spectra of $x$=0.0 and 0.11, as is evident in the left panel of Fig.~2.
In sharp contrast, the low temperature (T $<$ T$_t$) spectra of
$x$=0.0 and 0.11 clearly show a decrease of the DOS at E$_F$. Due to
technical reasons, the spectrum of NiS$_{0.89}$Se$_{0.11}$ was not
recorded at 25~K but at 80~K, which nevertheless, is also well below
T$_t$.  However, apart from the reduced DOS below T$_t$, all systems
are characterized by a finite DOS at E$_F$. It should be noted, that
significant changes are observed here in the PE spectra, which are
obviously related to the phase transition.  This means that the
surface sensitivity of the method does not appear to be a crucial
drawback in the particular case of these correlated systems.

        The above results clearly show that within the experimental
accuracy there is no gap between E$_F$ and the leading edge of the
spectrum of NiS at low temperature. However, this does not rule out
the possibility of an abrupt band edge as proposed in
Ref.~\cite{Nakamura1} with the Fermi level pinned very close~($<<$1~meV) 
to this edge.  Therefore, these results are not sufficient to
prove the ground state of NiS to be metallic with an absence of a gap
in the DOS. To establish that there is no gap directly above E$_F$,
the unoccupied DOS was probed using highly resolved PE spectroscopy
at various temperatures, progressively populating states above E$_F$
with increasing temperature. Since an insulator with a band gap
directly above E$_F$ cannot be populated, while a metal with
continuous and finite DOS above E$_F$ can be, the temperature
dependence of the spectra of a metal and an insulator are
fundamentally different. We show the near-E$_F$ spectra of NiS
collected at four different temperatures within the low temperature
phase in Fig.~3 (open symbols). At all temperatures, the spectra
cross a common energy point which, within experimental uncertainty,
is the Fermi energy. Furthermore, there is clear evidence of a
progressive and systematic development of spectral weight above E$_F$
and a corresponding depletion below E$_F$ with increasing
temperature. This is a convincing demonstration of finite DOS above
E$_F$ which is thermally populated. In order to put this on a
quantitative basis, we have carried out a least-squared-error
analysis of these spectra. It turns out that all four spectra can be
described in terms of a {\it single} DOS by only including the
Fermi-Dirac distribution for the respective temperatures (solid
lines). For comparison, Fig.~3 also displays the spectrum of NiS at
300~K, i.e.  well above the transition.  The inset of Fig.~3
summarizes the results of the present study, showing the extracted
DOS of NiS of both the high-temperature and the low-temperature
phases.  It is evident that both phases are metallic, with a smaller
DOS at E$_F$ in the low-temperature phase.  This decrease of the DOS
at E$_F$ is clearly correlated to the phase transition, since it is
only observed for NiS and NiS$_{0.89}$Se$_{0.11}$, while for
NiS$_{0.85}$Se$_{0.15}$ and NiS$_{0.87}$Se$_{0.17}$, which do not
undergo the phase transition, the DOS does not change (see Fig.~2).

        The results discussed so far establish that the ground state
of NiS and related compounds is definitely metallic with a large and
continuous DOS across E$_F$, and the increase in resistivity across
the phase transition is contributed by a decrease in the DOS at and
near E$_F$. For a possible explanation of this decrease, it is
interesting to note the similarity between the temperature-dependent
optical properties of NiS and chromium metal, which also reveals a
characteristic dip in reflectivity below the antiferromagnetic
transition temperature (see Fig.~6.  in~\cite{Barker1}). This dip in
the case of Cr has been convincingly demonstrated~\cite{Barker2} to
arise from the antiferromagnetic ordering, opening up gaps in the
electronic band structure only along specific directions in the
Brillouin zone, with the total DOS being continuous across E$_F$.
From these results, it appears reasonable to interpret the observed
decrease of the DOS at E$_F$ in NiS below T$_t$ (Figs.~2 and 3) as
well as the dip in the optical reflectivity~\cite{Barker1} as a
reflection of the opening of gaps along specific directions only,
caused by the antiferromagnetic ordering.  The present results do not
offer any clear explanation for the observed temperature independence
of the resistivity below the transition down to the lowest
temperatures.  This question is of course outside the scope of the
present work; however it is tempting to speculate on the mechanism of
such a metal-to-anomalous-metal transition. If a metallic state
exhibits a temperature-independent resistivity below a certain
temperature, it is evident that a temperature-independent scattering
dominates the transport properties, since there is no evidence for
further modification of the DOS with temperature within the low
temperature phase (see Fig.~3). The well-known mechanism of impurity
scattering, which gives rise to the constant resistivity of metals at
low temperatures ($<$~10~K), is unlikely to dominate the transport
properties up to temperatures as high as $\approx$~260~K. A rather
plausible mechanism, on the other hand, could be spin scattering, if
the antiferromagnetic order is incommensurate with the lattice. Since
the antiferromagnetic moments ($\approx$2.1~${\mu}_B$~\cite{Sparks2})
are fully developed at 260~K, with the extrapolated Neel temperature
being 1000~K, the spin scattering mechanism may possibly be
insensitive to temperature changes.  It is necessary, however, to
specifically address such issues both theoretically and
experimentally in order to establish whether such scattering
mechanisms could be independent of temperature over such a wide
temperature range and be dominant over other mechanisms to explain
the detailed temperature dependence of the transport properties
across the phase transition. In conclusion, the present results
establish that the first-order phase transition observed in NiS and
related compounds is a metal-to-anomalous-metal transition, instead
of being a metal-to-insulator transition, settling the age-old debate
on this issue.

\acknowledgments

        DDS thanks the Freie Universit\"at Berlin for hospitality
during part of this work. SRK thanks the CSIR, Government of India,
for financial support. Financial supports from Department of science
and Technology and Board of Research in Nuclear Sciences, Government
of India, are acknowledged. The work in Berlin was supported by the
Bundesminister f\"ur Bildung, Wissenschaft, Forschung und
Technologie, project Nos. 13N-6601/0 and 05 625 KEC.

\section{figure captions}

Fig.~1. Photoemission (PE) spectra of polycrystalline NiS,
NiS$_{0.83}$Se$_{0.17}$, and Ag in a narrow region around E$_F$
recorded at 25~K. Within experimental accuracy ($<$~1~meV), all
spectra show completely overlapping Fermi cut-offs.  The solid line
represents a fit of the Ag spectrum with an experimental resolution
of 9~meV (FWHM).

Fig.~2. PE spectra of NiS$_{1-x}$Se$_x$ with x=0.0, 0.11, 0.15 and
0.17, recorded above (left) and below (right) the transition
temperature T$_t$. For T$>$T$_t$ all samples show the same spectral
weight at  E$_F$, while for T$<$T$_t$ a substantial decrease is
observed for NiS and NiS$_{0.89}$Se$_{0.11}$. The insets display a
wider energy range, demonstrating the proper normalization of the
spectra.

Fig.~3. PE spectra of NiS recorded at various temperatures below the
phase transition temperature T$_t$ (open symbols), together with a
300 K spectrum. The inset displays the derived DOS for T$>$T$_t$ and
T$<$T$_t$.


\begin{references}
\bibitem{Sarma} Our transport measurements establish that the present samples
have very similar resistivities as reported for
single crystals~\cite{Koehler1}.
\end{references}
\end{document}